# One way to Lorentz's Transformations

E.G.Bessonov

*P.N.Lebedev Physical Institute RAS*

*There are many papers devoted to derivation of Lorentz Transformations (LT). Many people have proposed alternative derivations. Their analysis allows looking at LT and their consequences from different standpoints. At the same time it is important to choose the simplest and general way of derivation of LT for textbooks, to make them more accessible to a wide circle of laymen interested in the relativistic physics and to leave the W.Voight's, J.Larmor's, G.Lorentz's, A.Einstein's and other derivations for the history. A short, clear and simple derivation of generalized LT based on the Principle of Relativity and unknown dependence of the clocks tick rate (time dilation) on their velocity is presented in this paper.*

## I. Introduction

The laws of Nature are discovered by people in the process of carrying out experiments, the analysis of the received data and the search of the equations more precisely describing these data. If new experimental data aren't described well by earlier found equations, then other more exact equations are chosen. Such laws of mechanics as the Galileo Galilei Principle of Relativity (GPR, 1632), Galilean Transformations (GT, 1638) and three laws discovered by Isaac Newton (1687) in the analysis of experimental data received by his predecessors, became the basis of the well-composed, consistent theory called the theory of the Galilei-Newtonian classical mechanics. It seemed that foundations of this theory are unshakable. However the progress in studying the light and electrodynamics phenomena in the 19th century led to discovery of the Maxwell equations (1861) which turned out to be non-invariant relative to GT, i.e. changed their form under transition from one reference frame to another one. It meant that if Maxwell's equations are true then the GT have a limited area of applicability. Different ways were discussed to find new transformations. Among them:

1. GPR and GT are true. Maxwell's equations are incorrect.
2. In a reference frame connected with the ether the Maxwell equations are true, the velocity of light doesn't depend on the velocity of the source and the direction of propagation. In reference frames moving relative to ether the Maxwell equations change their form. The GPR and GT are broken.
3. The Principle of Relativity (Postulate) formulated by H.Poinkare in 1904 (PPR) is valid: «laws of the physical phenomena should be identical to the motionless observer and to the observer making uniform motion so we have no and we cannot have any way to define, whether there are we in similar motion or not» [1], [2] [1]. Pervasive ether is allowed.

The solution of the problem should be solved experimentally. The most significant and convincing results were received in the well-known negative Michelson - Morley experiments on detection of movement of the earth relative to ether [3], [4]. These experiments proceeded from 1881 to 1929. Ether remained imperceptible.

Various hypotheses for new experimental check had been put forward to search new transformations. Among them made a great impact on a course of search of LT hypothesis about the length contraction of moving objects (Fitzgerald, 1891, 1893); G. Lorentz, 1892) and the being in clouds hypothesis about change of the rate of the moving clocks tick were considered. The consequences from solutions of electrodynamics problems based on the use of the Maxwell equations and existing non-relativistic dynamics of an electron led to these hypotheses (compression of electromagnetic fields of homogeneously moving charged particles in the longitudinal direction, the wrong law of change of the energy of moving particle fields from their velocity, the dependence of the revolution time of the electron in the molecular on the molecular velocity).

---

[1] *On sense, the PPR and similar principal used by A. Einstein later (1905) doesn't differ from GPR: "in a cabin of the ship, moving homogeneously and without rolling, you won't find out neither on one of the surrounding phenomena, nor on something that will begin to occur with you, whether the ship moves or is motionless". They include electromagnetic and gravitational phenomena, which could include Galilei (his formulation is more general). All formulations set the equality of the inertial reference frames. Therefore we will use later only one term GPR.*

New transformations of coordinates and time should take the form

$$x = f_x(x',y',z',t',v), \quad y = f_y(x',y',z',t',v), \quad z = f_z(x',y',z',t',v), \quad t = f_t(x',y',z',t',v),$$

$$x' = f_x(x,y,z,t,-v), \quad y' = f_y(x,y,z,t,-v), \quad z' = f_z(x,y,z,t,-v), \quad t' = f_t(x,y,z,t,-v). \tag{1}$$

The last four equations in (1) follow from the first four equations, GPR and accounting of the velocity sign of one reference frame relative to another one.

It appeared that many solutions for functions $f_\alpha$ exist. Voight W. was the first who in 1887 included time $t'|_{v\neq 0} \neq t$ in transformations of coordinates. He received transformations, relative to which the wave equation for the free electromagnetic field was the invariant but his transformations were not in the agreement with the experiment. Scales on transverse axes in the contradiction to GPR changed proportional to the relativistic factor, the time scale changed as a square of the relativistic factor. The transformations which were in the agreement with the experiment were fond by J.Larmore in 1898 [6], G.Lorentz in 1899 (second approximation) and in 1904 (accurate calculation) [7]. By suggestion of A.Poiankare they were named by Lorentz Transformations (LT). The derivation of the LT was based on the invariance of Maxwell equations relative to these transformations [2]. G.Lorentz did not completely understand the sense of his transformations. A.Poiankare understood and estimated them deeply [8]. Earlier he already paid attention to the vital for the relativistic theory problem of synchronization of moving clocks, relativity of simultaneity and pointed to importance of the GPR for the inertial reference frames (presented some more detailed formulations for GPR) [3]. He interpreted LT as rotation in four-dimensional space-time, specified that they possessed group properties. Later on LT were derived by A. Einstein (1905) from GPR and from a principle of constancy of the light velocity (independence on $v$). His approach to the derivation was shorter and more graceful. It drew attention of many scientists. Besides, A. Einstein opened an interesting subject for laymen about paradox of clocks (twin paradox). In 1910 W.Ignatowsky derived generalized LT based on GPR, group axiomatic method using three inertial reference frames and without using of electrodynamics [10]. Next year Philippe Frank and Herman Roth published similar work in Annalen der Physik [11]. They developed W.Ignatowsky results and paid attention to the existence of more general fractionally linear transformations between two inertial frames. From their transformations the GT and LT followed as special cases. The derivation of the generalized LT in [10], [11] was based only on GPR which had not any quantitative data. Therefore entering into them a constant of integration, which had the dimension of velocity, remained uncertain. It was possible to define its value from experiments on the dependence of the lifetime of the exited states of molecules or the lifetime of unstable particles on their velocity. Checked experimentally Maxwell's equations could be used (they are invariant relative to the derived generalized LT if we suppose that entering into them and into Maxwell's equations constants are equal).

Thus the solution of the problem of search of the LT, the base of the new, consistent special theory of relativity, obliged by its emergency to a large number of outstanding mathematicians, physicists-theorists and experimenters was solved. In the present methodical paper a short, strict and, as it seems to us, simple for laymen derivation of the LT is presented. It is based on the GPR and one experimental point for time dilation. Other versions of derivation of LT are discussed.

---

[2] *The displacement current being entered by Maxwell at the time when all currents were considered closed, was considered as a guess. Later, when non-closed currents began to be considered, its introduction became the need following from the law of conservation of a charge. Experiments on generation of electromagnetic waves (1888) confirmed the need of its introduction.*

[3] *J. Larmor not only found the first, but in 1900 also consciously perceived the LT. So in the book [9] he calculated in the second order the dependence of a cycle period of an electron on the velocity of its molecule. He noted that in this case his orbit from the circular turns into the elliptic. The presented results were considered as reality, instead of as a trick (on G. Lorentz's terminology).*

## II. The derivation of G. Lorentz transformations based on the GPR and delay of the rate of the moving clocks tick.

Let's consider the inertial motionless reference frame $K$ and the frame $K'$ moving along the axis $x$ with a velocity $v$ (see Fig. 1). Axes $x, y, z$ of the frame $K$ and the corresponding axes of the frame $K'$ are equally directed, their origins at the moments $t = t' = 0$ coincide. Clocks in both frames are identical and synchronized. The GPR is valid. It means that the space is homogeneous, isotropic and communication between frames is described by linear equations.

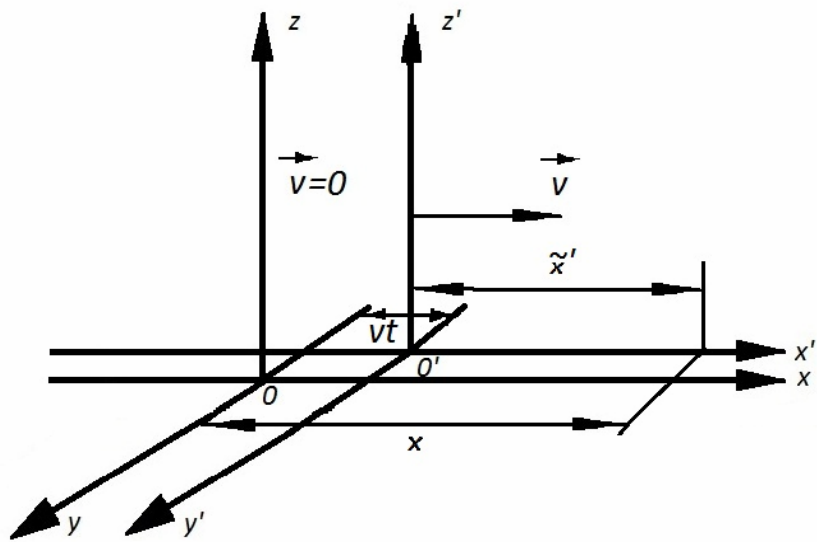

**Fig. 1**

In this case the scale rulers located in the transverse directions $y, z$ and $y', z'$ in the reference frames $K$ and $K'$ can be compared by their overlapping which is simultaneous for observers of each reference frame. That is why their dimensions, according to GPR, do not depend on the velocity. Hence it follows transformations for the transverse directions

$$y = y', \quad z = z'. \tag{2}$$

Let's take as a principle the experimental fact that the rate of moving clocks tick located in the reference frame $K'$ less $g(v)$ times than one for clocks at rest in the motionless frame $K$. Here, for the present, the dilation factor $g(v) \geq 1$ is the unknown function of the velocity $v$. It means that observers of the frame $K$ using their clocks located along the axis $x$, will find that moving past clocks, located at the origin of the reference frame $K'$, in a time interval $t$ will show time [4]

$$t' = t / g(v). \tag{3}$$

According to indications of these observers the clocks located at the origin of moving reference frame $K'$ for an interval of time $t$ will pass a way $l = vt$. The observers located in moving frame $K'$ will find that for appropriate time $t'$ the clocks located at the origin of the frame $K$, will pass a way $l' = vt'$ which, according to (3), is equal $l' = l / g(v)$, i.e. $g(v)$ times smaller. It means that objects located along the longitudinal axis in the frame $K$ for observers, located in the frame $K'$, look smaller $g(v)$ times. Scale ruler in this case is the part of $x$ axis located in $K$ frame between origins 0 and 0'. This ruler is observed from the frame $K'$. The length of the rule $l$ is determined by the time of going the clock located at the origin of the frame $K'$ past the ruler. The length of the rule $l'$ is determined by the time (2) of going the ruler past the origin of the frame $K'$ [12]. According to GPR (equality of frames $K$ and $K'$) the objects in the frame $K'$ being observed from the reference frame $K$, also look contracted the same number of times $g(v)$. Thus, from the law of time dilation and GPR other important law follows: in the inertial reference frame moving objects look contracted $g(v)$ times [5].

---

[4] *Time dilation of clocks takes place both for homogeneously moving clocks and for accelerated clocks moving, e.g., along circular orbits in accelerators.*

Coordinate $x$ and following from the law of the length contraction coordinate $x' = g(v)\tilde{x}'$ are related by the equation $\tilde{x}' = x - vt$ (see Fig. 1). Here the value $\tilde{x}'$ is the length of the section $x' = 0'x'$ measured in the motionless reference frame $K$. It follows the transformation
$$x' = g(v)(x - vt). \qquad (4)$$

According to GPR the similar transformation between coordinates $x$, $x'$, $t'$ follows for the measurements produced in the moving reference frame $K'$:
$$x = g(v)(x' + vt'). \qquad (5)$$
We took into account that in this case the velocity of the reference frame $K$ measured in the reference frame $K'$ has the opposite sign.

The system of equations (2), (4), (5) relates coordinates and time of the events taking place in the motionless and moving reference frames. In such way, it is possible to assert that the task of the relativistic transformations search is solved if the function $g(v)$ is found experimentally in the form of a table being approximated by some analytical function.

Having substituted the coordinate $x'$ defined by (4) in (5), we will find the explicit dependence of time in moving frame $K'$ on longitudinal coordinate and time in the frame $K$:
$$t' = g(v)\{t - [g^2(v) - 1]x / vg^2(v)\}. \qquad (6)$$

Having substituted the coordinate $x$ in the expression (4), we will find the explicit dependence of time in the motionless frame $K$ on longitudinal coordinate and time in the frame $K'$:
$$t = g(v)\{t' + [g^2(v) - 1]x' / vg^2(v)\}. \qquad (7)$$

The expressions (6), (7) can be presented in the form
$$t' = g(v)[t - xv / C^2(v)], \qquad (8)$$
$$t = g(v)[t' + x'v / C^2(v)], \qquad (9)$$
where the value $C^2(v) = v^2 g^2(v) / [g^2(v) - 1]$.

The transformation low of the longitudinal particle velocity $v_x'$ from the moving frame to motionless one, according to the expressions (5), (9) is
$$v_x = \frac{dx}{dt} = \frac{v_x' + v}{1 + v_x' v / C^2(v)}. \qquad (10)$$

Notice that according to non-electrodynamics experimental data (see e.g. [13]) the factor $g(v)$ with high accuracy $\sim 10^{-3}$ coincides with the relativistic factor $\gamma(v) = 1/\sqrt{1 - \beta^2}$, where $\beta = v/c$. It follows that the value $C(v)$ does not depend on $v$, has the velocity dimension and approximately equal to the light velocity $c$. Maxwell equations are invariant under the LT if we accept $C=c$. From this particular experimental electrodynamical data more accurate numerical value for the fundamental constant $C$ follow.

Thus, the system of the equations (2), (4) - (7) is a generalized LT (received from GPR only and including the fundamental unknown constant C). The dependence of the function $g(v)$ including in them can be taken from the experimental data being written down in the form of tables. They can be approximated by suitable analytical functions (for example $\gamma(v)$). Two of four equations (4) - (7) are independent. The equations for $x$, $t$ are direct and for $x'$, $t'$ are inverse generalized LT.

Explicit analytical dependence $g(v)$ can be found by different ways theoretically starting from GPR. The shortest way is related with the analysis of the expression (10). If to put that the velocity $v_x'$ represents the velocity of the third inertial reference frame $K''$ than, according to GPR, it should enter in (10) equally with the velocity $v$. That is why if we move from the $K$ system to $K''$ then the total velocity $v_x$ in (10) will not be changed (see [14][15]) but the universal constant in (10) will be a function of $v_x'$ ($C(v_x')$). It follows $C(v) = C(v_x')$ at arbitrary $v_x$ and $v$. Therefore $C(v)$ should be a constant independed on the velocity $v$: $C(v) = C$. From here and from the definition $C(v)$ the dependence follows $g(v) = 1/\sqrt{1-(v/C)^2}$.

---

[5] *It's true the converse: from the law of the length contraction of moving objects and GPR the law of the delay of the rate of a tick of moving clocks follows.*

Another way of search of the value $C(v)$ is connected with the process of synchronization of clocks by their transfer from one point of the frame $K'$ to another one at small velocity relative to this frame (see [16], page 24). For this purpose it is possible to carry along the axis $x'$ the clocks located in the origin of moving frame $K'$ at a factor $g(v)_\delta = g(v) + [\partial g(v)/\partial v]\Delta v$, where $\Delta v = v_\delta - v \ll v$, $v_\delta$ is the velocity of carried clocks in the frame $K$. In this case for the observers of the frame $K$ the rate of a tick of the clocks moving with a factor $g(v)_\delta$ will differ from the rate of a tick of clocks moving with a factor $g(v)$ and so according to (3), there will be a time shift of moving clocks relative to clocks being in the origin of the moving frame $K'$ on the value $\Delta t' = -\Delta t \Delta g(v)/g^2(v) = -\Delta t \Delta v [(\partial g(v)/\partial v)/g^2(v)] = -[\tilde{x}'/g^2(v)][\partial g(v)/\partial v]$, where $\tilde{x}' = \Delta t \Delta v = x'/g(v)$. Thus, indications of clocks in the moving reference frame will be defined by indications of the clocks (3) located in its origin with the shift $\Delta t'$: $t' = t/g(v) - [\tilde{x}'/g^2(v)] \cdot [\partial g(v)/\partial v]$. It follows the indications of clocks in the frame $K$

$$t = g(v)\{t' + [x'/g^3(v)][\partial g(v)/\partial v]\}. \qquad (11)$$

For the observers of the frame $K'$ the synchronization of clocks in the motionless frame, according to GPR, will lead to the similar dependence

$$t' = g(v)\{t - [x/g^3(v)][\partial g(v)/\partial v]\}. \qquad (12)$$

The equations (11), (12) can be used the same way as equations (6), (7). They are identical in spite of difference in their analytical form. From equality of the second terms in the expressions (6) and (12) the equation $[g^2(v) - 1]/vg^2(v) = [1/g^3(v)][\partial g(v)/\partial v]$ or $dg(v)/dv = g(v)[g^2(v) - 1]/v$ follow. Its solution is:

$$g(v) = 1/\sqrt{1-(v/C)^2}, \qquad (13)$$

where $C = const$ is the constant of integration. It follows that the value $C(v)$ included in (8), (9) do not depend on the velocity: $C(v) = C$.

From the equations (6) - (13) the direct and inverse transformations from one reference frame to another one follow:

$$t = g(v)(t' + vx'/C^2), \quad t' = g(v)(t - vx/C^2). \qquad (14)$$

Thus, being based only on GPR we found the generalized LT (2), (4), (5), (14) and analytical dependence of the being included into them the factor $g(v)$ outside the electrodynamics. The constant $C$ in these transformations is a fundamental constant. It is possible to determine it from the expression (13) using one experimental point for $v \neq 0$. We have received W.Ignatowsky's results using more simple physical way and confirmed once again that more general (non-electrodynamics) laws of the Nature are responsible for foundations of the relativistic physics.

## III. Discussion

GPR and unknown time dilation factor for moving clocks $g(v)$ were the bases of our search of the space-time transformations between inertial reference frames. The length contraction law $l' = l/g(v)$ was derived first using GPR. Then the linear equation united the coordinates $x$, $x' = g(v)\tilde{x}'$ and time $t$ was presented in the form: $\tilde{x}' = x - vt$, where $\tilde{x}'$ is the contracted length of the moving section $x' = O'x'$ being measured in the motionless reference frame $K$. From this and GPR the generalized LT in the form $x' = g(v)(x - vt)$, $x = g(v)(x' + vt')$ was found. This system of two equations relates space and time coordinates in both reference frames [6]. From these equations follow the explicit form of two linear dependent equations for time: $t' = g(v)[t - xv/C^2(v)]$, $t = g(v)[t' + x'v/C^2(v)]$, where the value $C^2(v) = v^2 g^2(v)/[g^2(v) - 1]$. This is the generalized LT with the new unknown values $g(v)$, $C(v)$. Then we have proved that the value $C(v)$ does not depend on the velocity $v$ and represents some fundamental constant $C$ which is equivalent to W.Ignatowsky constant $1/n$. Two simple ways of the proof were presented. If $C=const$ then from the definition of $C(v)$ the value $g(v) = 1/\sqrt{1 - (v/C)^2}$ follow. Thus, the received equations are the generalized LT with the unknown constant $C$ having the velocity dimension. The unknown constant $C$ can be found both from electrodynamics and non-electrodynamics experiments.

This simple, short, physical version of the generalized LT derivation was produced in the framework of the elementary mathematics and logic analysis. None concrete dynamics or electrodynamics lows (e.g., Maxwell equations) were used.

If we will proceed from the assumption that all lows of Nature follow to GPR then the conclusion will follow that the light, gravitation and other velocities in free space have the same value both in the reference frame at rest or in the state of homogeneous rectilinear movement. In this case it is possible to use clocks based on periodic oscillations of the light wave packets in the open resonator [7]. The period of oscillation of the wave packet in such light clocks at rest is equal $T_0 = 2l/c$, where $l$ is resonator length. If a clock moves with the velocity $v$, its axis is transverse to their velocity then the wave packet at the reference frame at rest propagates under the angle to the clocks velocity. Its longitudinal component of the velocity is equal to $v$ and the transverse one is decreased to $c\sqrt{1 - (v/c)^2}$. The period of oscillations of the wave packet is increased in accordance with the dilation factor $g(v) = \gamma(v) = 1/\sqrt{1 - (v/c)^2}$. If we would substituted the dilation factor in the expression (3), then the derivation of LT's would be shortened (will be limited by expressions (2) - (7)). But in this case we would deviate from the general case leading to the generalized LT with the fundamental constant $C$.

Note that according to GPR the rate of clocks tick (3) does not depend on their orientation in space, the design (based on spring, revolution, electricity or gravitation) or from belonging to animated or unanimated Nature.

If Maxwell did not discovered his equations, the constant $c$ taken from both the non light and aged light experiments in the electrodynamics (see [17], p. 814) could be used in the generalized LT at least by intuition.

### IV. Conclusion

1. The derivation of the generalized LT based on the GPR and unknown dependence of the clocks tick rate on their velocity is presented. This simple and short version of derivation was produced in the framework of the elementary mathematics and logic analysis. Concrete dynamics or electrodynamics lows (e.g., Maxwell equations) were not used. The development of the relativistic theory was started from the first principles.
2. The laws of Nature are governed by GPR. They are obliged to be invariant relative to LT.

---

[6] *These two equations can be considered as the generalized LT (received from GPR only and including the fundamental unknown constant C) if the time dilation factor $g(v)$ is obtained in the experiments and its data presented in the form of tables and analytical expressions approximating them.*

[7] *The elementary open resonator consists of two concave mirrors located at some distance from each other along some axis. It is used in laser equipment.*

3. We hope that presented way of derivation of LT can become more general, natural and clear for a wide circle of readers including laymen. It seems to us, that at this stage it is simpler to discuss existing questions and paradoxes, and some of them will disappear automatically [8].
4. It is interesting to note that concrete, complicated forms of analytical expressions for LT follow from GPR only. GPR does not carry definite numerical information obtained in the experiment (as a matter of fact GPR follow from a simple observation by Galileo Galilee of natural phenomena at Earth and from windows of the homogeneously moving ships). Only one additional experimental datum is necessary to find the value of their fundamental constant. One can only wonder at the fact that there are very many information in the GPR and its significant play in Nature.

Some of the considered subjects can be found at one or another interpretation in the papers [18] – [22] and cited references.

---

[8] *For example, we started with experimentally established law on time dilation for moving objects. Therefore the most difficult problem for perception with twins' paradox disappears as time dilation is an experimental fact.*